\documentclass[%
superscriptaddress,
preprint,
amsmath,amssymb,
aps,
pra,
]{revtex4-2}

\usepackage{graphicx}
\usepackage{dcolumn}
\usepackage{bm}

\begin{document}
	
\preprint{}
	
	\title{Hidden Variables: Rehabilitation of von Neumann's Analysis, and Pauli's Uncashable Check}%
	
	\author{R. Golub}
	\email{rgolub@ncsu.edu}
	\affiliation{Physics Department\\
		North Carolina State University, Raleigh, NC}
	\author{S.K. Lamoreaux}%
	\email{steve.lamoreaux@yale.edu}
	\affiliation{%
		Yale University, Department of Physics\\
		New Haven, CT
	}%
	
	\date{\today}
	
	\begin{abstract}
		
		In his book \textit{The Mathematical Foundations of Quantum Mechanics}, published in 1932, J. von Neumann performed an analysis of the consequences of introducing hidden parameters (hidden variables) into quantum mechanics.
		He arrived at two principal conclusions,  that hidden variables relating to spatial coordinates and momentum cannot be incorporated into the existing theory of quantum mechanics without major modifications, and that if they did exist, the theory would have already failed in situations where it has been successfully applied. This analysis has been taken as an ``incorrect proof" against the existence of hidden variables, possibly due to a mistranslation of the German word \textit{prufen}. von Neumann's so-called proof isn't even wrong as such a proof does not exist. One of the earliest attempts at a hidden variable theory by D. Bohm requires a major modification of quantum mechanics, which supports von Neumann's first principal conclusion. However, the Bohm theory has no experimental consequences, so W. Pauli referred to it as an ``uncashable check."  As there are no observable consequences, the Bohm theory is possibly a counterexample to von Neumann's second conclusion that hidden variables would have already led to a failure of the theory.
		To our knowledge, a successful hidden variable extension to quantum mechanics with testable consequences has not yet been produced, suggesting that von Neumann's analysis is worthy of rehabilitation, which we attempt to provide straightforwardly.

	\end{abstract}
	
	\maketitle
	
	
	\section{Background}
	
	{\it The Mathematical Foundations of Quantum Mechanics}, (MFQM) by J. von Neumman, published in 1932, is a masterpiece of theoretical physics.\cite{MFQM} In MFQM, von Neumann provided a complete exposition of the fundamentals of quantum mechanics as a linear operator theory, through Hermitean operators and Hilbert spaces.  von Neumann further applied mathematical analysis to the problems of quantum theory, such as quantum statistical mechanics and the measurement processes, that continue to serve as the basis of the theory.
	
	One of von Neumann's analyses in MFQM that has remained of central interest for nearly a century is the consequences of introducing hidden parameters (hidden variables) into quantum mechanics.
	He arrives at two principal conclusions, that hidden variables cannot be incorporated into the existing theory of quantum mechanics without major modifications to the fundamental theory, and that if they did exist, quantum mechanics would have already failed in situations where it has been successfully applied.

	One of the earliest attempts at a hidden variable theory was by D. Bohm and represents a major modification of quantum mechanics, and therefore is compatible with von Neumann's first principal conclusion. However, as Bohm's theory has no testable consequences, it is possibly a counterexample to von Neumann's second principal conclusion, that hidden variables would lead to observable deviations from the usual form of quantum mechanics.  Because the Bohm theory has no experimental consequences, W. Pauli referred to it as an ``uncashable check."  We might assume, wrongly or rightly, that von Neumann did not entertain the notion of a modification of quantum mechanics with no testable consequences.  
	
	To our knowledge, a successful hidden variable extension to quantum mechanics with testable consequences has not yet been produced, leading us to conclude that von Neumann's analysis is worthy of rehabilitation, which we attempt to provide straightforwardly.
	
	Our analysis follows a different tack than the recent work by Acu\~na \cite{acuna} who gives a detailed overview of the intense debate concerned with the interpretation and validity of von Neumann's work which has unfolded over several decades. (See, e.g., \cite{dieks,bub})  He points out that the Bohm theory abandons the notion that physical observables are the result of hermitian operators and therefore expands quantum theory beyond its Hilbert space formulation, while that formulation is the foundational basis of von Neumann's analysis and he does not offer the possibility of expanding beyond it. In our view, von Neumann does not provide a proof, but an investigation of the possible outcomes due to the introduction of hidden parameters within the Hilbert space formulation of quantum mechanics, as we will discuss.
	
	Dieks \cite{dieks} writes that ``According to what has become a standard history of quantum mechanics, in 1932 von Neumann persuaded the physics community that hidden variables are impossible as a matter of principle, after which leading proponents of the Copenhagen interpretation put the situation to good use by arguing that the completeness of quantum mechanics was undeniable. This state of affairs lasted, so the story continues, until Bell in 1966 exposed von Neumann's proof as obviously wrong."
	Much earlier, Grete Hermann criticized von Neumann's analysis and raised objections similar to Bell's, although her work was only later fully appreciated.\cite{grete1}
	
	In fact, in MFQM von Neumann showed that adding hidden parameters to the existing theory leads to a logical contradiction. His analysis is in essence a {\it reductio ad absurdum} argument, that if hidden variables exist, it would be possible to construct dispersion-free coordinate states, after showing that such states are not possible within the existing Hilbert space framework of quantum mechanics. He concluded that the present quantum theory would have already given false predictions if hidden variables existed. However, he left open the possibility of hidden parameters while recognizing that they would require a vastly modified theory.

	In this paper, we summarize von Neumann's discussion and suggest some new ways of looking at the situation.
	It is noted that the presentations in MFQM are sometimes rather opaque.
	
	A few mistranslations of the German original have possibly led to confusion. Specifically, on p. 321 bottom, ``Therefore $V=c'U$ ...." makes no sense, as it is an assumption. The German word {\it daher} should be translated not as ``therefore," but as ``because," noting that although this translation is (now) seldom used, it is correct in this context.
	
	A more troubling mistranslation occurs on
	p. 210 of MFQM, with the original being, ``Bis uns eine genauere Analyse der Aussagen der Quantenmechanik
	nicht in die Lage versetzen wird, die M\"oglichkeit der Einf\"urung
	verborgener Parameter objektiv zu pr\"ufen, was am oben erfolgen
	wird, wollen wir auf diese Erkl\"rungsm\"oglichkeit verzichten," for which the relevant part was translated to English as ``prove objectively the possibility of the introduction of hidden parameters." (See p.109 of the 1932 edition, \cite{MFQM},) Pr\"ufen does not mean prove, but should be translated as ``examine" or ``test," which is precisely what von Neumann does in MFQM.  (It should be noted that Hermann likely read the original German so it is difficult to assess the overall consequences of the mistranslations.)

	\section{von Neumann's analysis of hidden parameters}
	\subsection{Overview}
	von Neumann's analysis progresses in several steps and is centered on the density matrix. In MFQM, he placed quantum mechanics on a firm mathematical footing
	based on operators and vectors in Hilbert space and showed that it did not
	need to rely on the then mathematically dubious $\delta $-function introduced by Dirac. He also put considerable effort into the discussion of the
	possibility that the statistical behavior associated with the quantum states
	might be due to fluctuations in some unknown parameters (hidden variables), whose variations would lead to random behavior, just as averaging over the positions and velocities of the individual molecules leads to the statistical behavior in
	classical statistical mechanics.
	
	von Neumann considers quantum mechanics to be characterized by
	the relation
	for the expectation value
	of a physical variable represented by the
	Hermitian operator $\mathbf{R},$ in the state represented by the Hilbert
	space vector $\left\vert \phi \right\rangle $:%
	\begin{equation}
		\left\langle R\right\rangle =\left\langle \phi \left\vert \mathbf{R}%
		\right\vert \phi \right\rangle . \label{aa}
	\end{equation}%
	This is the inner product of the state $\left\vert \phi \right\rangle $ with
	the state $\mathbf{R}\left\vert \phi \right\rangle.$
	
	We expand $\left\vert \phi \right\rangle $ in a complete orthonormal set $%
	|\psi _{j}\rangle $:
	\begin{equation}
		\left\vert \phi \right\rangle =\sum_{j}\left\vert \psi _{j}\right\rangle
		\left\langle \psi _{j}\right\vert \left. \phi \right\rangle \,,
		\label{ch14_1a}
	\end{equation}%
	so that
	\begin{align}
		\left\langle \phi \left\vert \mathbf{R}\right\vert \phi \right\rangle &
		=\sum_{j,k}\left\langle \phi \right\vert \left. \psi _{j}\right\rangle
		\left\langle \psi _{j}\right\vert \mathbf{R}\left\vert \psi
		_{k}\right\rangle \left\langle \psi _{k}\right\vert \left. \phi
		\right\rangle \\
		& =\sum_{j,k}\left\langle \psi _{k}\right\vert \left. \phi \right\rangle
		\left\langle \phi \right\vert \left. \psi _{j}\right\rangle \left\langle
		\psi _{j}\right\vert \mathbf{R}\left\vert \psi _{k}\right\rangle \\
		& =\sum_{j,k}\left\langle \psi _{k}\right\vert \rho \left\vert \psi
		_{j}\right\rangle \left\langle \psi _{j}\right\vert \mathbf{R}\left\vert
		\psi _{k}\right\rangle =\sum_{j,k}\rho _{kj}R_{jk} \\
		& =\mathrm{Tr}\left( \rho \mathbf{R}\right)=\left\langle R\right\rangle \label{ch14_2a}
	\end{align}%
	where $\rho $ is the projection operator onto the state $\left\vert \phi \right\rangle $
	given by
	\begin{equation}
		\rho =\left\vert \phi \right\rangle \left\langle \phi \right\vert
		\label{ch14_2b}
	\end{equation}%
	The operator $\rho $ is called the density matrix
	or statistical operator and represents all the information available
	about the quantum state, equivalent to the wave function. It allows the
	introduction of statistical mixtures
	of different states.
	
	Equation (\ref{ch14_2b}) holds when the system is in a single quantum state,
	called a pure state. In the case when the system is described by a
	statistical ensemble consisting of systems in various states $\left\vert \phi
	_{i}\right\rangle $ with probabilities $w_{i},$ what is called a mixed
	state, the density operator is given by
	\begin{equation}
		\rho =\sum_{i}w_{i}\left\vert \phi _{i}\right\rangle \left\langle \phi
		_{i}\right\vert \,.
	\end{equation}
	
	If (\ref{aa}) is considered an assumption it can never be disproved, so von
	Neumann took one step back and replaced (\ref{aa}) by a set of assumptions
	which can lead to (\ref{ch14_2a}). Without specifying a definite method for
	calculating expectation values, he listed several properties he would expect
	to be features of such a method. Among these are%
	\begin{equation}
		\left\langle \alpha R\right\rangle =\alpha \left\langle R\right\rangle
		\label{ch14_4}
	\end{equation}%
	where $\alpha $ is a number and
	\begin{equation}
		\left\langle R+S+T+..\right\rangle =\left\langle R\right\rangle
		+\left\langle S\right\rangle +\left\langle T\right\rangle +...
		\label{ch14_3}
	\end{equation}%
	Von Neumann was well aware of the problem connected with the addition of
	operators that do not commute (see, e.g., MFQM p. 298). Equation (\ref{ch14_3}) is certainly valid in
	quantum mechanics since it follows from (\ref{ch14_2a}), but to cover the
	more general case, von Neumann defined the sum of the operators $R+S+T+...$
	to be that operator which had the expectation value given by the right-hand
	side of (\ref{ch14_3}) (i.e., an implicit definition). Indeed, he considered
	explicitly the case of the Hamiltonian for the electron in a hydrogen atom,
	in which
	\begin{equation}
		E=\langle H\rangle =\left\langle {%
			\frac{p^{2}}{2m}}-{\frac{e^{2}}{r}}\right\rangle =\left\langle {\frac{%
				(p_{x}^{2}+p_{y}^{2}+p_{z}^{2})}{2m}}\right\rangle -\left\langle {\frac{e^{2}%
			}{\sqrt{x^{2}+y^{2}+z^{2}}}}\right\rangle \label{ch14_5}
	\end{equation}
	
	Measurement of the first term on the RHS of (\ref{ch14_5}) requires a
	momentum measurement while that of the second term is a coordinate
	measurement.  The sum is measured in an entirely different way. Each
	measurement requires an entirely different apparatus. The variables $p_{i}$
	and $x_{i}$ (and hence functions of them) cannot be individually
	simultaneously precisely determined due to their noncommutativity.
	Nonetheless, as in the present case of the hydrogen atom, the sum of $%
	p^{2}/2m$ and $-1/r$ is defined and has a precise value, which is
	the energy. The role of a hidden variable theory is to allow the
	determination of the expectation values of $p$ and $r$, as can be done in
	classical mechanics, and provide an internal framework that corresponds to classical expectations. In such a theory, quantum mechanical uncertainty would arise simply due to our inability to see these hidden parameters or variables.
	
	von Neumann then proceeded to derive (\ref{ch14_2a}) from
	equation (\ref{ch14_3}) and several other assumptions as follows:
	
	Obviously (\ref{ch14_3}) follows from the trace rule (\ref{ch14_2a}) so that in any case where (\ref{ch14_3})
	fails, (\ref{ch14_2a}) and hence quantum mechanics will also not be valid.
	
	Starting with the representation of an operator as a matrix,
	\begin{equation}
		\mathbf{R}=\sum_{mn}\left\vert m\right\rangle \left\langle m\right\vert
		\mathbf{R}\left\vert n\right\rangle \left\langle n\right\vert
		=\sum_{mn}\left\vert m\right\rangle \left\langle n\right\vert R_{mn}\,,
	\end{equation}%
	von Neumann takes the expectation value
	of both sides (note that this applies to any
	method of taking expectation values that satisfies (\ref{ch14_4})) and (\ref%
	{ch14_3})):
	\begin{align}\left\langle \mathbf{R}\right\rangle & =\left\langle \sum_{mn}\left\vert
		m\right\rangle \left\langle n\right\vert R_{mn}\right\rangle
		=\sum_{mn}\left\langle \left\vert m\right\rangle \left\langle n\right\vert
		\right\rangle R_{mn} \\
		& =\sum_{mn}\rho _{nm}R_{mn}=\mathrm{Tr}\left( \rho \mathbf{R}\right)
		\label{trformula}
	\end{align}%
	where we defined $\rho _{nm}=\left\langle \left\vert m\right\rangle
	\left\langle n\right\vert \right\rangle $ the expectation value of the
	operator $\left\vert m\right\rangle \left\langle n\right\vert .$
	
	\subsection{Implications of hidden variables}
	
	If the statistical variations in experimental results were due to averaging
	over unknown ``hidden variables," the ensembles
	described by quantum states would consist of
	separate subensembles, each with some exact value of all physical variables.
	These values would have to be eigenvalues
	of the corresponding operators for the results
	to agree with observations, which are found to always yield eigenvalues, in
	agreement with the current version of quantum mechanics. In addition, it
	would be possible, at least in principle, to separate out these
	subensembles, so that each of the separate ensembles would have the property
	that all variables had exact values and there would be no scatter in the
	measured values of observables. Von Neumann called such states
	`dispersion-free' states, and the resulting subensembles he called
	`homogeneous ensembles'. A ``homogeneous ensemble" is an ensemble that can only be divided into
	subensembles that are identical to the original ensemble. If the statistical
	behavior associated with the quantum states was due to hidden variables,
	these quantum states would represent ``inhomogeneous ensembles."
	
	\subsection{``Dispersion-free" states and homogeneous ensembles in quantum
		mechanics}
	
	Starting with Equation (\ref{ch14_2a}), von Neumann wrote the dispersion \cite{albertson} (mean square fluctuation) of the variable represented by
	the operator $\mathbf{R}$ as
	\begin{equation}
		\sigma ^{2}=\mathrm{Tr}\left[ \rho (\mathbf{R}-\langle \mathbf{R}\rangle)^{2}\right] =\mathrm{Tr}%
		\left[ \rho (\mathbf{R}^{2}-\langle \mathbf{R}\rangle^{2})\right] =\langle \mathbf{R^{2}}%
		\rangle -\langle \mathbf{R}\rangle ^{2}
		\label{bg15}
	\end{equation}%
	because \(\langle \mathbf{R}\rangle\) is a number.
	\(\sigma^2\) is, in general, non-zero. It is zero if $\mathbf{R}|\phi \rangle
	=R|\phi \rangle $, which means that $|\phi \rangle $ is an eigenstate of $%
	\mathbf{R}$, and the system is in that eigenstate.
	
	We call a state ``dispersion-free" when from (\ref{bg15})
	\begin{equation}
		\mathrm{Tr}\left[ \rho \mathbf{R^{2}}\right] =\left( \mathrm{Tr}\left[ \rho
		\mathbf{R}\right] \right) ^{2} \label{dispfree}
	\end{equation}%
	for all Hermitian measurement operators
	$\mathbf{R}$.
	
	von Neumann then considered $\mathbf{R}$ to be the projection operator onto a state $|\phi\rangle$, so that $\mathbf{R}=\mathbf{%
		P}_{\phi}=|\phi\rangle\langle\phi|$. When $\mathbf{P}_{\phi}$ operates on \textit{any} state $|\Psi\rangle$ that is formed from a complete set of eigenfunctions that include $|\phi\rangle$,
	\begin{equation}
		\left( \mathbf{P}_{\phi} \right) ^{N}|\Psi\rangle=\mathbf{P}%
		_{\phi}|\Psi\rangle=c_{\phi}|\phi\rangle\, ,
	\end{equation}
	which means that the same result is obtained when $\mathbf{P}_{\phi}$ is
	applied multiple times.
	
	With $\mathbf{R}=\mathbf{P}_{\phi }$ in Eq. (\ref{dispfree}), (note: \(\mathrm{Tr}\left[ \rho\boldsymbol{P}_{\left[ \phi\right]
	}^{{}}\right]\)=\(\langle \phi |\rho |\phi \rangle\))
	\begin{equation}
		\mathrm{Tr}\left[ \rho\boldsymbol{R}^{2}\right] =\left( \mathrm{Tr}\left[ \rho
		\boldsymbol{R}\right] \right) ^{2}=\mathrm{Tr}\left[ \rho\boldsymbol{P}_{\left[
			\phi\right] }^{2}\right] =\mathrm{Tr}\left[ \rho\boldsymbol{P}_{\left[ \phi\right]
		}^{{}}\right] =\left( \mathrm{Tr}\left[ \rho\boldsymbol{P}_{\left[ \phi\right]
		}^{{}}\right] \right) ^{2}%
	\end{equation}
	\begin{equation}
		\langle \phi |\rho |\phi \rangle =\langle \phi |\rho |\phi \rangle ^{2}
	\end{equation}%
	for all $|\phi \rangle $. Therefore,
	\begin{equation}
		\langle \phi |\rho |\phi \rangle =0\ \ \mathrm{or}\ \ \ \langle \phi |\rho
		|\phi \rangle =1\,.
	\end{equation}%
	This should hold for any normalized state $|\phi \rangle $. Taking $\phi
	=c_{1}|\phi _{1}\rangle +c_{2}|\phi _{2}\rangle $ we vary $c_{1}$ and $c_{2}
	$ in a continuous manner, so that $|\phi \rangle $ starts at $|\phi
	_{1}\rangle $ and ends at $|\phi _{2}\rangle $ (MFQM p. 320-321). During
	this process, the relation $\langle \phi |\rho |\phi \rangle =0$ or $1$ must
	hold over the entire variation, and we therefore conclude that
	\begin{equation}
		\rho =0\ \ \mathrm{or}\ \ \rho =1
	\end{equation}%
	for a dispersion-free ensemble. This means that $\rho $ is a diagonal
	matrix with elements all 0 or all 1. The case of all zero diagonal elements is trivial because in this case $R=0$ always, which is not possible in a realistic system.
	
	The obvious dispersion-free case of $\rho$ being in a pure state of the $%
	\mathbf{R}$ basis is possible for one or even for several measurements
	operators $\mathbf{R}$, but it cannot simultaneously be true for \textit{all}
	measurement operators since some of them do not commute. Thus, such a state
	would not be dispersion-free because it would still show dispersion in the
	measurements of some observables.
	
	In the case of $\rho =1$, von Neumann tells us (top of MFQM, page 321) that there is a normalization problem, in that
	\begin{equation}
		\mathrm{Tr}\left( \rho \right) =N\rightarrow \infty
	\end{equation}%
	where $N$ is the number of elements (dimension of the state space). However, this sum should be unity if $\rho $ represents the system average density matrix. This is because the diagonal matrix elements are the probabilities of being in each eigenstate, and
	\begin{equation}
		\sum_{n}w_{n}=1=\mathrm{Tr}\left( \rho \right) .
	\end{equation}%
	This would seem to preclude the possibility of constructing a
	general density matrix with $\rho=1$ because it obviously cannot be normalized.
	On the other hand, we read (MFQM, p. 310, where Exp[\(\mathbf{R}\)] is the expectation value):
	\begin{quote}
		... we shall admit not only \(R=\mathrm{Exp}(\mathbf{R})\) functions representing expectation values, but also functions corresponding to relative values -- i.e. we allow the normalization condition (\({\rm Exp}(1)=1\)) to be dropped.
		....(\({\rm Exp}(1)=\infty\)) is an entirely different matter and it is actually for this sake that we want this extension...
	\end{quote}
	and (MFQM, p. 320, and \(U=\rho\) in our notation) that ``It is for an infinite \(\textrm{Tr}(U)\) only that we have essentially relative probabilities and expectation values." So it seems infinite traces are not \textit{a priori} forbidden.
	This together with the fact that $\rho =1$ is not dispersion-free, while it is supposed to be a solution for zero dispersion, raises some questions about the relevance of this solution. If instead of considering the dispersion for an observable represented by a projection operator, we focus on a general operator with $\textbf{R}^2\neq\textbf{R}$ the only solution to (\ref{dispfree}) valid for all \textbf{R} is $\rho=0$.
	
	Thus, von Neumann concluded that if the assumptions Eqs. (\ref{ch14_4}) and (\ref%
	{ch14_3}) hold, dispersion-free states do not exist.
	It is evident that the impossibility of dispersion-free states is related to the Heisenberg uncertainty principle. Non-commuting observables cannot be dispersion-free in the same state and the commutation relations prescribe a limit to the possible accuracy of their measurement.
	
	\subsection{No hidden variables ``theorem"}
	After the investigation of the possibility of dispersion-free states von Neumann went on to study the possibilities of uniform or homogeneous ensembles, that is ensembles that can be divided into sub-ensembles which would all give the same expectation value for every physical quantity, i.e all physical quantities would have the same probability distribution in every sub-ensemble. He then shows that such ensembles exist and they are pure quantum states and hence not dispersion-free. As this argument does not seem to have been challenged we have placed it in an Appendix. If the dispersion shown in the homogeneous ensembles (quantum states) was due to some hidden variables with different values it would be possible to separate them into sub-ensembles according to the different values of these hidden variables in contradiction to their property of being homogeneous.
	Either of the two results would rule out hidden variables coexisting with quantum mechanics. Dispersion-free
	states are impossible and the quantum states are homogeneous ensembles so
	it is impossible, according to quantum mechanics, to break up a quantum mechanical ensemble into subensembles with different physical properties. According to von Neumann:
	\begin{quote}
		But this [the existence of hidden variables] is impossible for two reasons:
		First, because then the homogeneous ensemble in question could be
		represented as a mixture of two different ensembles, contrary to its
		definition. Second, because the dispersion-free ensembles, which would have
		to correspond to the ``actual" states (i.e., which consist only of systems in
		their own ``actual" states), do not exist. (%
		MFQM, p. 324.)
	\end{quote}
	von Neumann summarizes his argument:
	
	\begin{quote}
		It should be noted that we need not go any further into the mechanism of the
		``hidden parameters" since we now know that the
		established results of quantum mechanics can never be rederived with their
		help. In fact, we have even ascertained that it is impossible that the same
		physical quantities exist with the same functional connections, if other
		variables (i.e., \textquotedblleft hidden parameters\textquotedblright )
		should exist in addition to the wave functions. Nor would it help if there
		existed other, as yet undiscovered, physical quantities, in addition to
		those represented by the operators in quantum mechanics, because the
		relations assumed by quantum mechanics (i.e., \textbf{I.}, \textbf{II.})\cite{relations} would have to fail already
		for the currently known quantities, those that we discussed above. It is
		therefore not, as is often assumed, a question of a re-interpretation of
		quantum mechanics,---the present system of quantum mechanics would have to
		be objectively false, in order that another description of the elementary
		processes than the statistical one be possible. (MFQM, p. 324.)
	\end{quote}
	
	\section{Reactions to von Neumann's discussion of hidden variables}
	von Neumann's discussion of hidden variables unleashed a vigorous debate that has lasted almost a hundred years. A reasonably comprehensive summary has been given by Acu\~na.\cite{acuna} A year after the publication of MFQM, Grete Hermann, a philosophy student who was defending the
	philosophical tradition that causality was a necessary constituent for any scientific view of the
	world, produced a criticism of von Neumann's argument (\cite{grete1}).
	She'd had extensive discussions with Heisenberg, and von Weizs\"acker (for details see \cite{dieks}).
	Hermann called attention to the fact that the assumption
	\[\mathrm{Exp}[a\mathbf{R}+b\mathbf{S}]=a\mathrm{Exp}[\mathbf{R}]+b\mathrm{Exp}[\mathbf{S}]\]
	fails for quantities that are not simultaneously measurable (their operators do not commute). She then noted that von Neumann had recognized this fact and implicitly defined the operator for the sum of physical quantities as that operator whose expectation value was the sum of the expectation values on the rhs. She deduces from this that von Neumann needs another ``proof for quantum mechanics". Then she claimed that he finds this in the use of
	\[R=\mathrm{Exp}[\mathbf{R}]=\langle\phi|\mathbf{R}|\phi\rangle\]
	for the expectation value of \textbf{R} in the state \(|\phi\rangle\). For this definition, the expectation value of a sum is given by the sum of the expectation values. However, this only holds for those ensembles whose definition depends only on those physical quantities which are considered in the present quantum theory, those quantities which determine the state \(|\phi\rangle\). She states that von Neumann's argument for the necessity of dispersion applies only to such ensembles. It has not been shown that the expectation value has the above form for ensembles that agree not only in the state \(|\phi\rangle\) but also in possibly yet to be discovered, presently unknown, quantities (hidden variables). A physicist who only knows a given system by its Schr\"odinger wave function is bound to find themself limited by the uncertainty principle. For the von Neumann proof to hold we must assume that \(\langle\phi|\mathbf{R}|\phi\rangle\) represents the average value of measurements in any ensemble whose elements agree with each other not only with respect to \(|\phi\rangle\) but also with respect to arbitrary as yet undiscovered quantities. Hermann stated
	\begin{quote}
		That all these ensembles have the same average value is an assumption justified neither by previous experience nor by the hitherto confirmed theory of quantum mechanics. Without it, the proof of indeterminism collapses.
	\end{quote}
	
	This argument was repeated in a slightly different form in her ``Natural-Philosophical Foundations of Quantum Mechanics" whose English translation has been published in chapter 15 of \cite{grete1}.
	
	This work remained largely unknown and the von Neumann proof was generally accepted as providing strong support for the Copenhagen interpretation of quantum mechanics until in 1952 David Bohm (\cite{bohm}) reinvented a theory that had been proposed by de Broglie in 1927. For details of these theories see (\cite{bgskl}.) The Bohm theory treats the position of a particle as an exactly knowable hidden variable, but determining this with more accuracy than allowed by quantum mechanics requires an initial state that violates the uncertainty principle. Because of this Pauli called it a ``check that can't be cashed." \cite{pauli} 
	
	Bohm's theory led to an outbreak of papers analyzing and criticizing von Neumann's ``proof."  The most influential of these was a series of papers by John Bell. His reaction to reading Bohm's paper was as follows:
	\begin{quote}
		But in 1952 I saw the impossible done. It was in papers by David Bohm. Bohm
		showed explicitly how parameters could indeed be introduced into
		nonrelativistic wave mechanics, with the help of which the indeterministic
		description could be transformed into a deterministic one. More importantly,
		in my opinion, the subjectivity of the orthodox version, the necessary
		reference to the ``observer" could be eliminated.\cite{bellpilot} This [assumption (\ref{ch14_3})] is true for
		quantum mechanical states, \emph{it is required by von Neumann of the
			hypothetical dispersion free states also }[emphasis added]. At first sight,
		the required additivity of expectation values seems very reasonable, and it
		is the non-additivity of allowed values (eigenvalues) which requires
		explanation. Of course, the explanation is well known: A measurement of a sum
		of noncommuting variables cannot be made by combining trivially the results
		of separate operations on the two terms---it requires a quite distinct
		experiment. For example the measurement of $\sigma _{x}$ for a magnetic
		particle might be made with a suitably oriented Stern-Gerlach magnet. The measurement of $\sigma _{y}$ would require a different orientation and the
		measurement of $\left( \sigma _{x}+\sigma _{y}\right) $ a third and different orientation. But this explanation of the non-additivity of allowed values also established the non-triviality of the additivity of expectation
		values.\textquotedblright
	\end{quote}
	In another paper, Bell writes
	\begin{quote}
		The latter (\ref{ch14_3}) is a quite peculiar property of quantum mechanical
		states, not to be expected $a\ priori$. There is no reason to demand it
		individually of the hypothetical dispersion-free states, whose function it
		is to reproduce the \emph{measurable} properties of quantum mechanics when
		\emph{averaged over}. [emphasis added] \cite{hvbell}
	\end{quote}
	
	David Mermin\cite{mermin}
	characterized the assumption (\ref{ch14_3}) as ``silly" and quoted Bell in a
	published interview:
	\begin{quote}
		Yet the von Neumann proof, if you actually come to grips with it, falls
		apart in your hands! There is \emph{nothing} to it. It's not just flawed,
		it's \emph{silly}! .~.~. When you translate [his assumptions] into terms of
		physical disposition, they're nonsense. The proof of von Neumann is not
		merely false but \textit{foolish}! \cite{omni}
	\end{quote}
	
	Bell's point, quoted above with emphasis, is correct: In dispersion-free states, the expectation value of a variable would have to be one of the eigenvalues of the corresponding operator, and so
	equation (\ref{ch14_3}) could not apply. But contrary to Bell's point above,
	von Neumann's argument does not require (\ref{ch14_3}) to be true for dispersion-free
	states.
	To reiterate, von Neumann's logical argument, which has the appearance of circularity
	(sort of a \textit{reductio ad absurdum}) is based on the following three propositions:
	
	\begin{itemize}
		\item[\textbf{A}.] The sum of expectation values assumption, Eq. (\ref{ch14_3}).
		
		\item[\textbf{B}.] $\langle \mathbf{R}\rangle =\mathrm{Tr}(\rho \mathbf{R})$, Eq. (\ref{trformula})
		that is, the trace of $\rho $ times an operator gives the expectation value of
		that operator.,
		
		\item[\textbf{C}.] Dispersion-free states do NOT exist.
	\end{itemize}
	
	von Neumann has shown that \textbf{A}\ $\Rightarrow$\ \textbf{B}\ $\Rightarrow$\ \textbf{C}.
	If this conclusion is taken as the argument against hidden variables, then Bell, Hermann, and others are correct because \textbf{A} does not hold for
	dispersion-free states, and it is, therefore, a circular, silly argument.\cite{dieks}
	However, let us look at the logic.
	It is obvious that \textbf{B}$\Rightarrow$ \textbf{A} so that if \textbf{A} were false (which is the case for a dispersion-free state) \textbf{B} would have to be false i.e. the predictions of quantum mechanics would have to fail for already known quantities as von Neumann stated. On the other hand assumption \textbf{A} is only necessary for deriving \textbf{B}. But nobody would deny that \textbf{B} represents the essence of quantum mechanics. Since \textbf{B}\ $\Rightarrow$\ \textbf{C}, if \textbf{C} were false, i.e. there were dispersion-free states, \textbf{B} would have to be false and again the conclusion is that quantum mechanics would have to fail with the existence of hidden parameters.
	
	\section{conclusion}
	We have shown that examination of the logic of von Neumann's argument leads to the conclusion that the existence of hidden variables capable of allowing the exact prediction of all physical quantities would mean that quantum mechanics in its present form would have to be false, i.e. the existence of hidden variables would contradict quantum mechanics, and their inclusion requires a vastly modified theory. Of course, this follows already from the fact that physical quantities represented by non-commuting operators must satisfy an uncertainty relation.
	Another powerful argument against hidden variables has been presented by Pauli. In a letter to Fierz (Jan. 6, 1952) he wrote\cite{pauli2}:
	\begin{quote}
		I want to call special attention to the thermodynamics of ensembles, consisting of the same type of subensembles (Einstein-Bose or Fermi-Dirac statistics). What is important to me is not the energy values but the statistical weights, further the indifference of the thermodynamic-statistical reasoning to the "wave-particle" alternative and Gibbs' point that \textit{identical} or only similar states behave qualitatively differently. If hidden parameters exist, not only on paper, but determine a really different behavior of different single systems (e.g.particles) - according to their "real" values - so must - completely independent of the question of the technical measurability of the parameters - the Einstein-Bose or Fermi-Dirac statistics be completely disrupted. Since there is no basis to assume that the thermodynamic weights should be determined by only half (or a part of) "reality". Either two states are identical or not (there is no "similar") and if the $\psi$ function is not a complete description of single systems, states with the same $\psi$ function will not be identical. Every argument with the goal of saving the Einstein-Bose and Fermi-Dirac statistics from the causal parameter mythology must fail because it - taking into account the usual theory in which the $\psi$ function is a complete description of a state - declares the other half of reality to be unreal.
	\end{quote}
	In the case of Bohm's theory, each particle follows its own trajectory, and the wavefunction is in 3$N$ dimensional space and is even or odd under particle exchange. The hidden variable is the initial position of each particle, hence is not an additional degree of freedom because in ordinary quantum mechanics the particle positions are the degrees of freedom.  For identical particles, when the initial wavefunction is specified, the even or odd superposition results from the creation of a multiparticle state at each initial position. Overall, there is no surprise that the system evolution is indistinguishable from the usual formulation of quantum mechanics.

	\section{Appendix}
	
	von Neumann asked if there is a possibility for the existence of
	subensembles of the density matrix
	where each has a different but exact dispersion-free
	expectation values for a set of physical parameters, leading to dispersion
	of the full ensemble. The idea is that such subensembles could be chosen to produce a dispersion-free value of some hidden parameter and that the
	dispersion of the full ensemble is due to the spread in the exact values
	among the subensembles. These subensembles would be homogeneous ensembles. He
	then proves that any homogeneous ensemble is a quantum state
	and any quantum state is a homogeneous ensemble. This
	means quantum states cannot be divided into subensembles with different
	properties.
	
	Consider dividing the density matrix into two parts such that (using modern notation per \cite{albertson})
	\begin{equation}
		\rho=\rho_{1}+\rho_{2}
	\end{equation}
	$\rho$ would constitute a homogeneous ensemble if $\rho_{1}=c_{1}\rho$ and $%
	\rho_{2}=c_{2}\rho.$ Von Neumann proceeded by assuming we have a $\rho$ with
	such properties.
	
	Then he wrote $\rho_{1,2}$ as the following Hermitian operators:
	\begin{equation}
		\rho_{1}={%
			\frac{\rho|f_{0}\rangle\langle f_{0}|\rho}{\langle f_{0}|\rho |f_{0}\rangle}}
		\label{a1a}
	\end{equation}
	and
	\begin{equation}
		\rho_{2}=\rho-{\frac{\rho|f_{0}\rangle\langle f_{0}|\rho}{\langle
				f_{0}|\rho|f_{0}\rangle}}
	\end{equation}
	where $|f_{0}\rangle$ is a state vector such that $\rho|f_{0}\rangle\neq0 $.
	Because $|f_{0}\rangle$ is arbitrary, with only the constraint that its
	projection by $\rho$ is nonzero, this decomposition is general. Also, it is
	obvious that $\rho=\rho_{1}+\rho_{2}$ as required by the definition.
	
	For any state $|\psi \rangle $, the probability that the system is in that
	state is
	\begin{equation}
		w_{\psi }=\langle \psi |\rho |\psi \rangle \geq 0
	\end{equation}%
	for all $|\psi \rangle $, which follows from the definition of $\rho $. This
	means that $\rho$ is a \textit{positive definite operator}. Therefore
	\begin{equation}
		\langle f|\rho _{1}|f\rangle ={\frac{|\langle f|\rho |f_{0}\rangle |^{2}}{%
				\langle f_{0}|\rho |f_{0}\rangle }}\geq 0,
	\end{equation}%
	which shows that $\rho _{1}$ is positive definite. From a theorem proved on
	p. 101 of MFQM, we know that
	\begin{equation}
		|\langle n|\mathbf{R}|m\rangle |^{2}\leq \langle n|\mathbf{R}|n\rangle
		\langle m|\mathbf{R}|m\rangle \,, \label{vnproof}
	\end{equation}%
	so we immediately see that
	\begin{equation}
		\langle f|\rho _{2}|f\rangle ={\frac{\langle f|\rho |f\rangle \langle
				f_{0}|\rho |f_{0}\rangle -|\langle f|\rho |f_{0}\rangle |^{2}}{\langle
				f_{0}|\rho |f_{0}\rangle }}\geq 0\,.
	\end{equation}%
	Thus, both $\rho _{1}$ and $\rho _{2}$ are Hermitian (by construction) and
	positive definite, which also means that the trace is greater than zero.
	
	Now define a state
	\begin{equation}
		|\phi \rangle =%
		\frac{\rho |f_{0}\rangle }{K}
	\end{equation}%
	where $K$ is chosen so that $\langle |\phi \rangle =1.$ Since we have
	assumed $\rho $ represents a homogeneous ensemble, $\rho _{1}=c_{1}\rho $.
	Then
	\begin{align}
		\rho |f\rangle & =\frac{1}{c_{1}}\rho _{1}|f\rangle =\frac{1}{c_{1}}{\frac{%
				\rho |f_{0}\rangle \langle f_{0}|\rho |f\rangle }{\langle f_{0}|\rho
				|f_{0}\rangle }=\frac{K}{c_{1}}}|\phi \rangle \frac{\langle f_{0}|\rho
			|f\rangle }{\langle f_{0}|\rho |f_{0}\rangle }=\frac{{K}^{2}/c_{1}}{\langle
			f_{0}|\rho |f_{0}\rangle }|\phi \rangle \langle \phi |f\rangle \\
		& =C|\phi \rangle \langle \phi |f\rangle
	\end{align}%
	Thus
	\begin{equation}
		\rho =C|\phi \rangle \langle \phi |=C\mathbf{P}_{\phi }\,,
	\end{equation}%
	and we see that if $\rho $ represents a homogeneous ensemble then $\rho =%
	\mathbf{P}_{\phi }$ and represents a quantum state. Every homogeneous
	ensemble is a quantum state.
	
	Next, we turn around and assume that we have a system in a quantum state $%
	\left( \rho =\mathbf{P}_{\phi }\right) .$ Von Neumann then proved that a
	quantum state is necessarily a homogeneous ensemble. Starting with a state $%
	|f\rangle $ orthogonal to $|\phi \rangle ,$ i.e., $\langle f|\phi \rangle =0$
	so that $\rho |f\rangle =\mathbf{P}_{\phi }|f\rangle =|\phi \rangle \langle
	\phi |f\rangle =0$. With $\rho =\rho _{1}+\rho _{2}$, where $\rho _{1,2}$
	are both positive definite, we obtain the inequality
	\begin{equation}
		0\leq \langle f|\rho _{1}|f\rangle \leq \langle f|\rho _{1}|f\rangle
		+\langle f|\rho _{2}|f\rangle =\langle f|\rho |f\rangle =0\,,
	\end{equation}%
	so that $\langle f|\rho _{1}|f\rangle =0$. By considering Eq.~(\ref{vnproof}%
	), replacing $|n\rangle $ by $|g\rangle $, $|m\rangle $ by $|f\rangle $ and $%
	\mathbf{R}$ by $\rho _{1}$ we find $\langle g|\rho _{1}|f\rangle =0$ for any
	state $|g\rangle $. Since $|f\rangle $ is any state orthogonal to $|\phi
	\rangle $ and every $|f\rangle $ is also seen to be orthogonal to $\rho
	_{1}|g\rangle $, we must have%
	\begin{equation}
		\rho _{1}|g\rangle =c_{g}|\phi \rangle
	\end{equation}%
	with $c_{g}$ a constant \textit{that can depend on} $|g\rangle .$ Taking $%
	|g\rangle =|\phi \rangle $ yields%
	\begin{equation}
		\rho _{1}|\phi \rangle =c^{\prime }|\phi \rangle
	\end{equation}%
	Any state $|h\rangle $ can be written as
	\begin{equation}
		|h\rangle =|\phi \rangle \langle \phi |h\rangle +|h^{\prime }\rangle
	\end{equation}%
	where $|h^{\prime }\rangle $ is orthogonal to $|\phi \rangle .$ Then%
	\begin{align}
		\rho _{1}|h\rangle & =\rho _{1}|\phi \rangle \langle \phi |h\rangle +\rho
		_{1}|h^{\prime }\rangle \\
		& =c^{\prime }|\phi \rangle \langle \phi |h\rangle =c^{\prime }\mathbf{P}%
		_{\phi }|h\rangle
	\end{align}%
	where $\rho _{1}|h^{\prime }\rangle =0$ because $h^{\prime }$ is orthogonal
	to $|\phi \rangle $. Therefore, $\rho _{1}=c^{\prime }\rho $ and, because $%
	\rho _{1}+\rho _{2}=\rho $, then $\rho _{2}=\left( 1-c^{\prime }\right) \rho
	$ and any quantum state is a homogeneous ensemble.
	
	If hidden parameters existed, we could pick specific values of those
	parameters and form subensembles such that a measurement of that subensemble would produce exact values for some parameters or reduce
	the dispersion of some parameters. von Neumann's point with this proof was to show that quantum ensembles are not dispersion-free simply because they are composed of distinct subensembles, each of which has a different value of some hidden parameter, with the statistical nature of the full ensemble being due to scatter from the different values associated with each subensemble.
	
	To reiterate, there are two reasons for this. First, there are no
	dispersion- or scatter-free ensembles because for such $\rho =0$. Second, if
	we were to postulate that dispersion results from an ensemble representing a
	collection of subensembles which each have a specific value for some hidden
	parameter, that postulate would fail because the subensembles would be
	homogeneous ensembles which themselves are quantum states and hence not
	dispersion-free. As we have seen, the original ensemble, representing a
	quantum state, is a homogeneous ensemble, precluding its breakup into
	different subensembles.
	
	It is understood that all these results hold only in the case that quantum
	mechanics is valid.
	
	\bibliographystyle{unsrt}

\end{document}